\documentclass[
  aps,
  prb,
  twocolumn,
  superscriptaddress,
  floatfix
]{revtex4-1}

\usepackage[T1]{fontenc}
\usepackage{graphicx}
\usepackage{amsmath,amssymb}
\usepackage{bm}
\usepackage{xurl}
\usepackage[hidelinks]{hyperref}
\usepackage[final]{microtype} 

\usepackage{color}
\usepackage{hyperref}
\usepackage{listings}
\hypersetup{
    unicode=false,          
    pdftoolbar=true,        
    pdfmenubar=true,        
    pdffitwindow=false,     
    pdfstartview={FitH},    
    pdfauthor={Kevin Multani},     
    colorlinks=true,       
    linkcolor=blue,          
    citecolor=red,        
}
\graphicspath{{figures/}}

\begin{document}

\definecolor{dkgreen}{rgb}{0,0.6,0}
\definecolor{gray}{rgb}{0.5,0.5,0.5}
\definecolor{mauve}{rgb}{0.58,0,0.82}

\lstset{frame=tb,
  	language=Matlab,
  	aboveskip=3mm,
  	belowskip=3mm,
  	showstringspaces=false,
  	columns=flexible,
  	basicstyle={\small\ttfamily},
  	numbers=none,
  	numberstyle=\tiny\color{gray},
 	keywordstyle=\color{blue},
	commentstyle=\color{dkgreen},
  	stringstyle=\color{mauve},
  	breaklines=true,
  	breakatwhitespace=true
  	tabsize=3
}

\title{Phonon-induced electronic degeneracy breaking: a  SSAdNDP interpretation}
\author{Javiera Cabezas-Escares$^{*}$}
\author{Andrea Echeverri}
\author{Francisco Muñoz}
\affiliation{Departamento de Física, Facultad de Ciencias, Universidad de Chile, \& CEDENNA Santiago 7800024, Chile.}
\author{Anastassia N. Alexandrova$^{*}$}
\affiliation{Department of Chemistry and Biochemistry, University of California, Los Angeles, California 90095, USA}
\affiliation{ Center for Quantum Science and Engineering, University of California, Los Angeles, California 90095, USA}
\date{\today}

\begin{abstract}
This work explores how phonon perturbations can induce the breaking of electronic degeneracies near the Fermi level and how this response can be interpreted from a chemical perspective through the SSAdNDP method. We apply this approach to a family of structurally similar yet electronically distinct hexagonal materials—MgB$_2$, graphene, and hBN—to analyze how a single phonon mode simultaneously modifies the electronic structure (band dispersion) and the nature of chemical bonding (natural occupations and nodal patterns) in real space. Our results show that band splitting becomes physically relevant only when it is accompanied by an electronic redistribution, reflected in changes of the occupation numbers or bonding topology. Thus, SSAdNDP provides a direct bridge between reciprocal- and real-space representations, translating phenomena such as electron–phonon coupling into chemically intuitive reorganizations of multicenter bonds, and offering a unified framework to interpret vibrationally driven electronic effects in solids.
\end{abstract}
\maketitle
Since the formulation of Bloch's theorem \cite{bloch_uber_1929}, reciprocal space has served as the fundamental domain for describing the electronic structure of periodic systems, particularly within the physics community. However, while this approach—relying on band structures, Fermi surfaces, and k-points—is computationally efficient, it often lacks an intuitive chemical representation. Consequently, many phenomena possess a rigorous physical description but lack a complementary understanding based on local chemical bonding \cite{alexandrova_divide-and-conquer_2017}. This disconnect has motivated the use of bonding models to bridge the gap between electronic structure and material properties, enabling a more intuitive characterization of the system \cite{alexandrova_divide-and-conquer_2017}.\\ 
In this context, the Solid State Adaptive  Natural Density Partitioning (SSAdNDP) \cite{galeev_solid_2013} offers a conceptual and theoretical framework to \textit{translate} the reciprocal space information to real space. This transformation provides an electronic description by bonds and natural occupation (ON).\\
Some of the phenomena frequently studied in reciprocal space, and not in real space, are band splitting. Band splitting is a response to symmetry breaking, often explained in terms of the Jahn–Teller effect, typical for many transition‑metal oxides, chalcogenides, \cite{Jahn-Teller, PhysRevResearch.1.033131, Khomskii2020Orbital, PhysRevB.103.L121114, PhysRevMaterials.6.075004} or the Peierls transition, principally in a low dimensional matterials \cite{bozin2019local, PhysRevB.103.L121114, x6zl-t5hs}. In such cases, the system lowers its energy by adopting a new structural configuration, which in turn determines its magnetic \cite{bozin2019local,PhysRevB.103.L121114}, optical\cite{PhysRevB.68.045105}, conductive \cite{PhysRevMaterials.6.075004} among others \cite{Khomskii2020Orbital, PhysRevResearch.1.033131}. In the present work, we mimic the effect of dynamical perturbations—specifically, selected phonon modes—using the frozen phonon approach \cite{PhysRevB.26.3259, PhysRevB.29.1575}. Although these configurations do not correspond to new equilibrium geometries, the role of phonons in modifying the electronic structure is of broad relevance. Electron-phonon interactions are known to drive or strongly influence phenomena such as conventional superconductivity \cite{bardeen1957theory} and band topology or topological phase transitions \cite{garate2013phonon, antonius2016temperature}. \\
Understanding these effects from both reciprocal -and real-space perspectives can therefore provide valuable insight into the microscopic mechanisms behind technologically relevant quantum phases. In this work, we will show the evolution of the chemical bonds under perturbations. It allows chemical interpretation of the breaking of degenerated electronic states close to the Fermi energy $(E_f)$. For this proposal, we will describe three geometrically similar systems, but electronically different systems: MgB$_2$, graphene, and monolayer hBN. They are a superconductor, semiconductor, and insulator, respectively. While hBN does have electrons close to the Fermi level, it is instructive studying it. The above schema allows us to study the same phononic mode for every system to elucidate the differences and the correlation between the band splitting and the changes in the bond pattern.

\section{\label{sec:one}Systems and Computational Methods}

\subsection{\label{sec:oneCond}Studied Materials}
\begin{figure}[t]
    \centering
\includegraphics[width=0.9\linewidth]{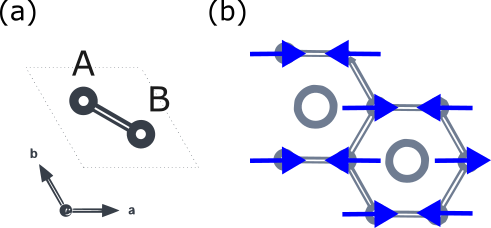}
    \caption{(a) General unit cell with A and B atoms. (b) $E_{2g}$ phonon mode in a hexagonal lattice}
    \label{fig:com_sys}
\end{figure}
We focus on three hexagonal systems that are structurally similar yet electronically distinct: MgB$_2$, monolayer hBN, and graphene. All share a unit cell with two atoms (A and B) arranged in a hexagonal lattice (See Fig. \ref{fig:com_sys}(a)). In graphene, the A and B atoms are non-equivalent carbon sites; in hBN, they correspond to chemically distinct boron and nitrogen atoms; and in MgB$_2$, the boron sublattice also forms an A–B network, but each hexagon is additionally centered by a magnesium atom. Despite these structural differences, the three systems nominally contain the same number of valence electrons per unit cell. However, this does not guarantee similar band structures or bonding patterns, which are instead determined by the chemical identity of the constituent atoms and their orbital interactions.
Additionally, the hexagonal symmetry of these systems gives rise to a common family of phonon modes. Among them, we focus on the in-plane bond-stretching $E_{2g}$ mode (See Fig. \ref{fig:com_sys}(b)), which is particularly relevant due to its strong electron–phonon coupling in MgB$_2$-like structures \cite{kong2001electron, yildirim2001giant,bohnen2001phonon,singh2022high}. This mode distorts the hexagonal lattice by modulating the bond lengths within the basal plane. Studying the same phonon mode across structurally similar systems, therefore, provides a consistent framework to reveal how this type of perturbation modifies chemical bonding patterns and to contrast these changes with the corresponding electronic band structures.

\subsection{\label{sec:methods}Computational Methods}
All calculations were performed using density functional theory as implemented in the Vienna Ab initio Simulation Package (VASP)\cite{vasp1,vasp2,vasp3,vasp4}. The Projector Augmented Wave (PAW) method \cite{paw} was used with the Perdew-Burke-Ernzerhof (PBE) exchange-correlation functional\cite{pbe}. The electronic wave functions were expanded in a plane-wave basis set with an energy cutoff of 500 eV for MgB$_2$ and graphene, 400 eV for hBN, which was deemed sufficient for convergence.\\
Self-consistent field (SCF) calculations were performed for the optimized structures of MgB$_2$, monolayer hBN, and graphene. A Gamma-centered k-point mesh was used for the Brillouin zone sampling, with grid sizes of $32\times32\times24$, $15\times15\times1$, and $25\times25\times1$ for MgB$_2$, hBN, and graphene, respectively. Electronic convergence was achieved when the total energy change between consecutive SCF steps was less than $10^{-8}$ eV\\
Phonon properties were calculated using the finite-displacement method as implemented in the Phonopy package\cite{togo2015first,togo2023first,togo2023implementation}, which interacts with VASP. We constructed a  $2\times2\times 1$ supercell for all materials to determine the interatomic force constants. 
To study the phonon mode $E_{2g}$, we selected the atomic displacements as a perturbation mode using the Phonopy modulation tool. The largest displacement magnitudes range from 0.024 to 0.032 {\AA} to keep it in the harmonic regimen.\\  The bands and SSAdNDP calculations are performed using the SCF calculations without relaxing these modes. Band structures are plotted using Pyprocar \cite{pyprocar,LANG2024109063} \\ 
The chemical bonding analysis was performed using the Solid State Adaptive Natural Density Partitioning (SSAdNDP) method \cite{galeev_solid_2013}, a technique developed for analyzing chemical bonding in periodic systems. This method was implemented using the software from the same research group. The analysis used the charge densities calculated from our VASP SCF runs. For the natural projections, we included the pob-TZVP atomic basis set \cite{peintinger2013consistent}, which is specifically designed for periodic calculations. The output files were then visualized using the VESTA software\cite{vesta}.

\subsection{\label{sec:analysis}Analysis Methods} 
To bridge the gap between the delocalized plane-wave basis sets used in our DFT calculations and a localized bonding picture, we employed the SSAdNDP method~\cite{galeev_solid_2013} combined with the periodic NBO projection scheme developed by Dunnington and Schmidt~\cite{dunnington_generalization_2012}.\\
In short, this procedure begins with the the density matrix in the atomic orbital basis, $P^{k,\mathrm{AO}}$. This matrix is transformed into the natural atomic orbital representation, $P^{k,\mathrm{NAO}}$, following the standard NBO procedure~\cite{reed1985natural} adapted for periodic boundary conditions~\cite{dunnington_generalization_2012}. Then, an inverse Fourier transform yields the real-space density matrix $P^{0s}_{\mu\nu}$, where $\mu$ and $\nu$ denote orbitals on atoms in the reference (0) and translated unit cells $(s)$, respectively. Interactions between unit cells are truncated at $s_{\mathrm{max}}$, such that $P^{0s_{\mathrm{max}}}_{\mu\nu} \approx 0$, while translational symmetry ensures $P^{0s} = P^{h(h+s)}$. \\
The AdNDP \cite{zubarev2008developing} procedure is then applied by constructing subblocks $P(ij\ldots k)$ in the NAO basis, which are analyzed to identify localized and multicenter bonds.  
\begingroup
\setlength{\arraycolsep}{0pt} 
\renewcommand{\arraystretch}{1.00}
\begin{equation}
P(ij\ldots k)=
\left[
\begin{array}{cccc}
P^{00,\mathrm{NAO}}_{ii} & P^{0s,\mathrm{NAO}}_{ij} & \cdots & P^{0t,\mathrm{NAO}}_{ik} \\
P^{0(-s),\mathrm{NAO}}_{ji} & P^{00,\mathrm{NAO}}_{jj} & \cdots & P^{0(t-s),\mathrm{NAO}}_{jk} \\
\vdots & \vdots & \ddots & \vdots \\
P^{0(-t),\mathrm{NAO}}_{ki} & P^{0(s-t),\mathrm{NAO}}_{kj} & \cdots & P^{00,\mathrm{NAO}}_{kk}
\end{array}
\right]
\end{equation}
\endgroup

After diagonalization of $P(ij\ldots k)$, the resulting eigenvectors are used to identify bonds and their occupations $N^{(ij\ldots k)}_{l}$, which are then subtracted to progressively deplete the density matrix, i.e.,  
\begin{equation}
\widetilde{P}^{0r,\mathrm{NAO}}_{ij}
= P^{0r,\mathrm{NAO}}_{ij}
- N^{(ij\ldots k)}_{l} \, c_{l,i} \, c^{*}_{l,j} \,
|n^{i}_{l}\rangle \langle n^{j}_{l}|.
\end{equation}
Here, $P^{0r,\mathrm{NAO}}_{ij}$ is the original density-matrix block in the NAO basis and $\widetilde{P}^{0r,\mathrm{NAO}}_{ij}$ is the corresponding block after subtraction of the identified $n$-center bond. The eigenvalue $N^{(ij\ldots k)}_{l}$ gives the bond occupation, while $c_{l,i}$ and $c^{*}_{l,j}$ are the eigenvector coefficients describing the contributions of atoms $i$ and $j$. This depletion procedure removes the density associated with the localized bond, preventing its double counting in subsequent multicenter searches.\\
In this way, the density matrix is decomposed into localized $n$-center bonds (two-center, three-center, etc.) characterized by natural bond occupations \cite{galeev_solid_2013,zubarev2008developing}. The method prioritizes the classical Lewis picture of electron pairing, in which the ideal limit corresponds to ON = 2 \cite{weinhold_valency_2005}. High electron pairing across multiple centers is interpreted as a delocalized bonding configuration.  \\
In this work, the analysis began with the identification of localized two-center bonds ($2c-2e$). These motifs were observed in all three systems (MgB$_2$, graphene, and h-BN), representing the localized $\sigma$-bonding framework connecting adjacent atoms in the hexagonal rings. This preliminary characterization is consistent with previous SSAdNDP studies on MgB$_2$ \cite{galeev_solid_2013} and in the NDP study of graphene \cite{popov_is_2012}, which established the dominance of the localized $\sigma$-skeleton. However, these localized bonds are not the central focus of our study. Instead, we concentrate on the distribution of the delocalized electron density.\\
To characterize the remaining valence density beyond this localized framework, we subsequently tested higher-order multicenter configurations to maximize the ON. For the primary system, MgB$_2$, two descriptions were established: a six-center bond ($n=6c$) confined to the boron rings, and a mixed eight-center ($n=8c$) configuration incorporating magnesium atoms (detailed in Appendix A, and both showed in the previous study \cite{galeev_solid_2013}). To ensure a rigorous comparison with isostructural systems, representative bonding motifs were selected based on the intrinsic chemical nature of each material. For graphene, we selected the $n=6c$ motif, corresponding to the covalent aromatic sextets shown in the references\cite{popov_is_2012}. In contrast, for hexagonal boron nitride (h-BN), the search algorithm identified a nitrogen-centered four-center bond ($n=4c$) as the optimal representation, reflecting the ionic localization of charge on the nitrogen sublattice. \\
We associate occupations close to two with states deeply buried below the Fermi level, which gradually become unoccupied as the bands approach the Fermi energy. The procedure was terminated once the six-center configurations were completely unoccupied.
Bonding patterns were visualized through isosurface representations, enabling us to monitor the emergence, disappearance, or reorganization of multicenter bonds under phonon perturbations.

\section{\label{sec:results} Results and Discussion}
\subsection{\label{sec:result_mgb2}MgB\texorpdfstring{$_2$}{2}: phonon-induced redistribution of electronic density}

MgB$_2$ is among the superconductors with the highest critical temperature ($T_c=39$ K) under ambient pressure \cite{kang2001mgb2, PhysRevLett.86.4656}. Its superconductivity is well described by the conventional Bardeen-Cooper-Schrieffer (BCS) theory \cite{bardeen1957theory}, in which electron–phonon coupling plays a central role. In particular, certain phonon modes are known to strongly modify the electronic structure and thereby promote Cooper pair formation. Among them, the in-plane bond-stretching mode that distorts the hexagonal boron lattice has been identified as the key driver of superconductivity in MgB$_2$ \cite{kong2001electron,  yildirim2001giant, bohnen2001phonon}. Motivated by this, in the present work, we perturb the system with this phonon mode (denoted $E_{2g}$ by its symmetry) and analyze the resulting changes in both the band structure and the real-space bonding patterns.

The natural occupations (ON) for the multicenter bonds were obtained from the SSAdNDP analysis. The hexagonal boron network, stabilized by the inclusion of magnesium, can be described by two main bonding motifs: six-center (6c) B–B bonds and mixed eight-center (8c) configurations that incorporate Mg. The presence of Mg slightly increases the ON values, as its charge donation partially fills the boron-derived bands near the Fermi level. However, Mg does not contribute directly to the bonding network; its primary role is to electronically stabilize the boron layer \cite{PhysRevLett.86.4656}, thereby reducing the extent of charge redistribution under phonon perturbation. This observation aligns with recent studies using bonding-based descriptors like integrated crystal orbital bonding index (iCOBI) , which suggest that symmetric and electronically saturated bonding environments exhibit enhanced resistance to anharmonic perturbations \cite{belli2025chemical}. For this reason, the comparative analysis between frozen-phonon and equilibrium configurations focuses on the 6c bonding patterns within the boron sublattice, where phonon-induced rearrangements are most clearly manifested. The complete analysis, including the 8c bonds with Mg, is provided in Appendix \ref{sec:Aone}.

\begin{figure}[t]
    \centering
    \includegraphics[width=0.95\linewidth]{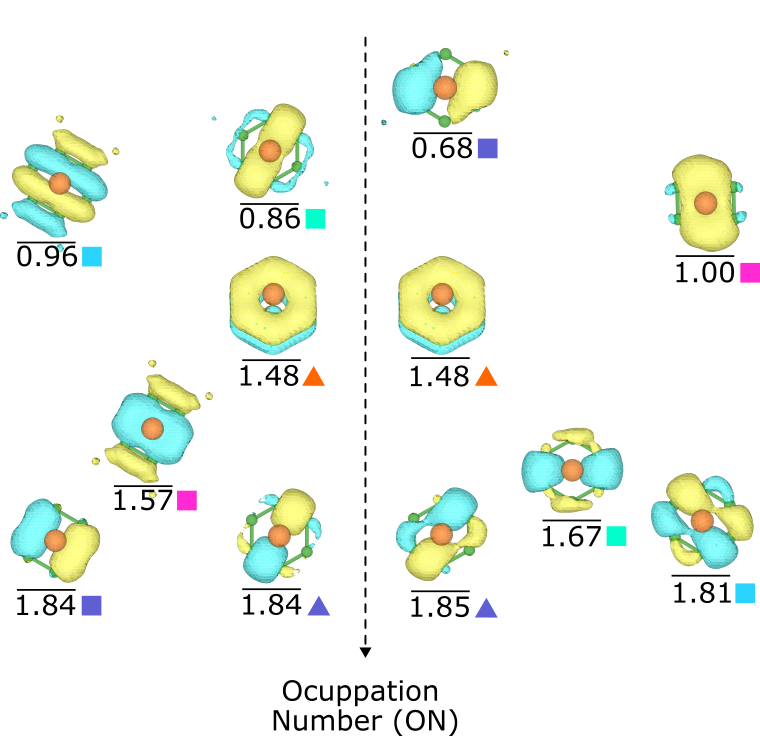}
    \caption{MgB$_2$: Bonding configurations and their occupation numbers (ON) for the 6c case. The left side displays the unperturbed system, while the right side shows the system perturbed by an in-lattice $E_{2g}$ phonon. The color-coding links correspond to bond geometries in each scenario; a square represents a change in the ON, and triangles show bonds that remain stable.
}
    \label{fig:bond_mgb2}
\end{figure}
\begin{figure}[t]
    \centering
    \includegraphics[width=\linewidth]{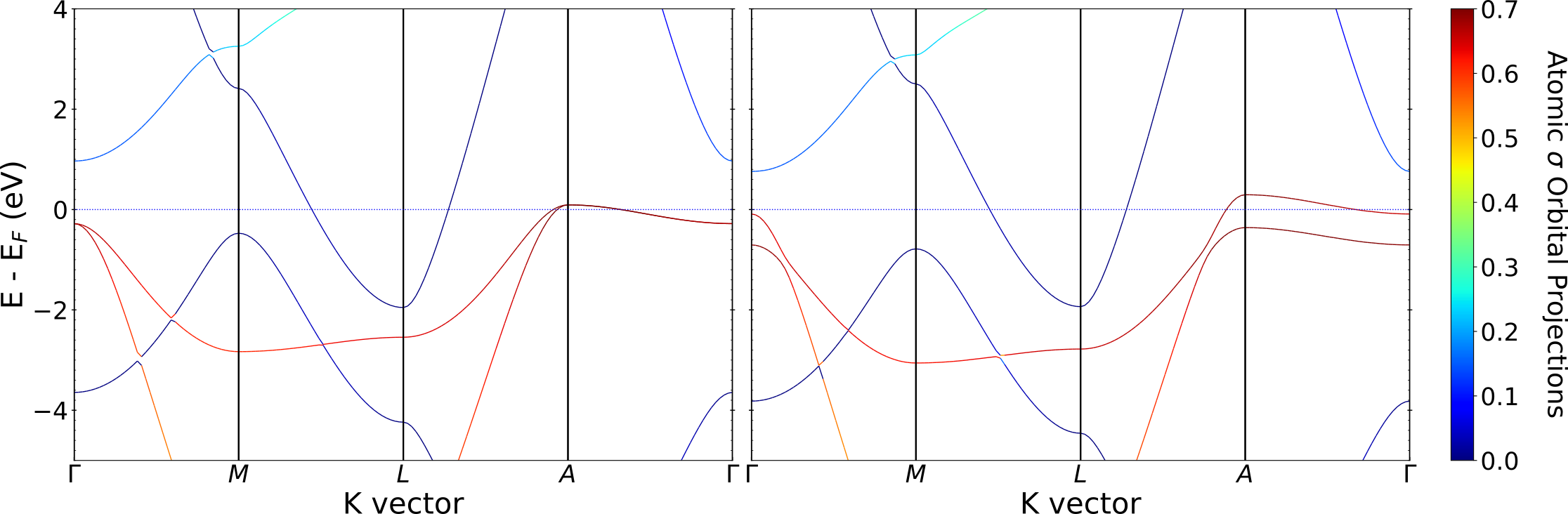}
    \caption{MgB$_2$: Comparison of the band structure for the (left) unperturbed and (right) perturbed system. The colors indicate the projections onto the $s, p_x$, and $p_y$ orbitals associated with the $\sigma$ bonds, while blue denotes the unique contribution from the $p_z$ orbitals.}
    \label{fig:bands_MgB2}
\end{figure}

The lattice deformation induced by the $E_{2g}$ phonon leads to a clear modification of the electronic structure (see Fig.\ref{fig:bands_MgB2}): the degenerate $\sigma$ bands are split, and the corresponding 6c bonds exhibit noticeable changes in both shape and occupation number. In contrast, the multicenter $\pi$ bonds (orange triangle in Figure \ref{fig:bond_mgb2}) remain remarkably robust, preserving both their topology and ON values under distortion. In the unperturbed system, two degenerate $\sigma$ bonds display occupations close to 1.8 |e| (purple square and triangle), two intermediate delocalized bonds show ON $\approx$1.5 |e| (pink square and orange triangle), and two $\sigma$ weaker bonds have ON $\approx$ 0.9 |e| (cyan and green squares). After introducing the frozen phonon, those $\sigma$ degeneracies are lifted: the occupations redistribute significantly (e.g., 1.84 $\to$ 0.68 |e|. 0.86 $\to$ 1.67 and 0.96 $\to$ 1.81), following the same trend as the $\sigma$ bands derived from boron  $s, p_x, p_y$ orbitals in Figure \ref{fig:bands_MgB2}. Meanwhile, the $\pi$-type multicenter bonds, associated with B $p_z$ orbitals, remain unchanged—mirroring the stability of the corresponding blue $\pi$ bands in the band structure.

This joint analysis of electronic bands and real-space bonds shows that the $E_{2g}$ phonon not only splits $\sigma$-band degeneracies but also redistributes electronic density within the boron layer, leading to a genuine reorganization of the bonding network. This behavior contrasts sharply with the response observed in graphene and hBN, where the same phonon mode lifts degeneracies without producing significant changes in the bonding topology or occupation numbers.

\subsection{\label{sec:hbn-graph}Graphene and hBN: lifting of degeneracies without bonding reorganization}

Due to their similar symmetry, hBN, graphene, and MgB$_2$ exhibit the same phonon mode in the plane $E_{2g}$ previously analyzed. However, the response of each system to this perturbation differs markedly in its electronic and bonding characteristics. While the $E_{2g}$distortion deforms the hexagonal lattice and modifies the local bond geometry, its effect on the electronic bands near the Fermi level is minor, and therefore does not alter the overall electronic behavior of these systems. The SSAdNDP analysis confirms this picture: the natural bond occupations (ON) and the number of nodal planes remain essentially unchanged, indicating that no real reorganization of bonding occurs.

In graphene, the band structure contains a degenerate state at the $\Gamma$ point approximately 3 eV below the Fermi level, and a Dirac cone at the $K$ point crossing the Fermi level. Although the $E_{2g}$ mode slightly lifts these degeneracies—opening a small gap at $K$ and splitting the lower-lying bands—the Fermi surface remains unaffected. 

Correspondingly, the SSAdNDP representation shows only internal rearrangements among bonds with the same number of nodes and similar ON values, implying that the energy landscape and bonding preferences remain unaltered. This behavior is consistent with the aromatic stability previously characterized for $\pi$ bonds \cite{popov_is_2012}, where the bonding pattern is shown to be robust against local variations.

\begin{figure}[t]
    \centering
    \includegraphics[width=0.85\linewidth]{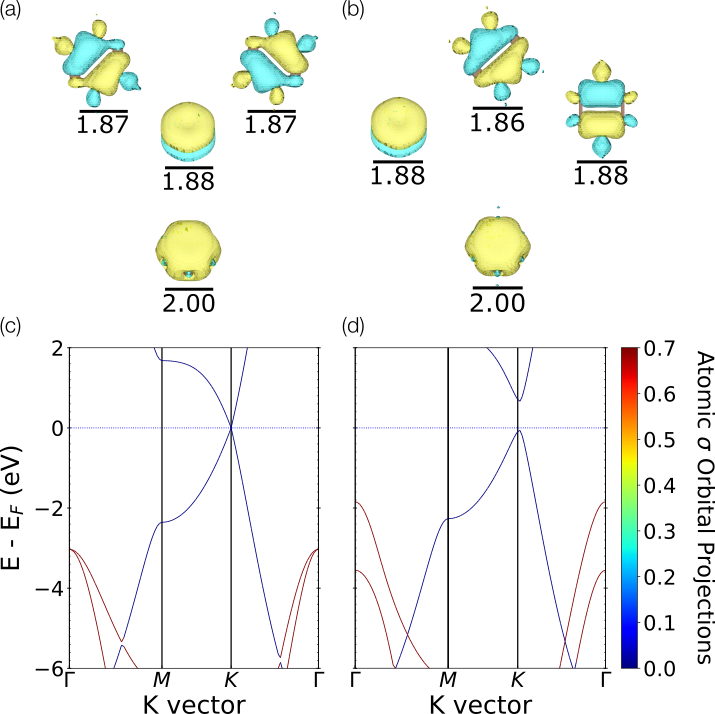}
    \caption{ Bonding analysis and electronic structure of graphene under the $E_{2g}$ phonon perturbation. Top panels display the SSAdNDP bonding motifs for the (a) unperturbed equilibrium state and (b) the distorted structure. The orbitals are sorted by ON, revealing a topologically invariant network where bonding preferences remain unaltered. Bottom panels present the electronic band structure for the (c) unperturbed and (d) perturbed systems. The color gradient represents the atomic projection onto the in-plane $\sigma$ orbitals ($s, p_x, p_y$)}
    \label{fig:graphene}
\end{figure}

Analogous to graphene, hexagonal boron nitride (h-BN) exhibits remarkable topological robustness under phonon perturbation. However, the underlying bonding mechanism differs significantly; the B–N bonds possess a polar and partially ionic character \cite{roy2021structure, muramatsu2003angle, ooi2005electronic}. Consequently, h-BN is an electrical insulator with a wide bandgap at the $K$ point, standing in contrast to the Dirac cone feature of graphene. Under the same $E_{2g}$ phonon perturbation, while the degenerate bands at the $\Gamma$ point undergo splitting, the electronic structure near the Fermi level remains largely unaffected.

Electron Localization Function (ELF) analysis confirms the strong ionic nature of the B–N interaction \cite{ooi2005electronic}, showing electron density—including the $\pi$ orbital contributions—heavily localized around the nitrogen atoms. Topologically, the ELF reveals triangular contours within the hexagonal rings directed towards the boron sites, yet the maximum localization remains firmly anchored within the nitrogen sublattice \cite{ooi2005electronic}.

\begin{figure}[t]
    \centering
    \includegraphics[width=0.85\linewidth]{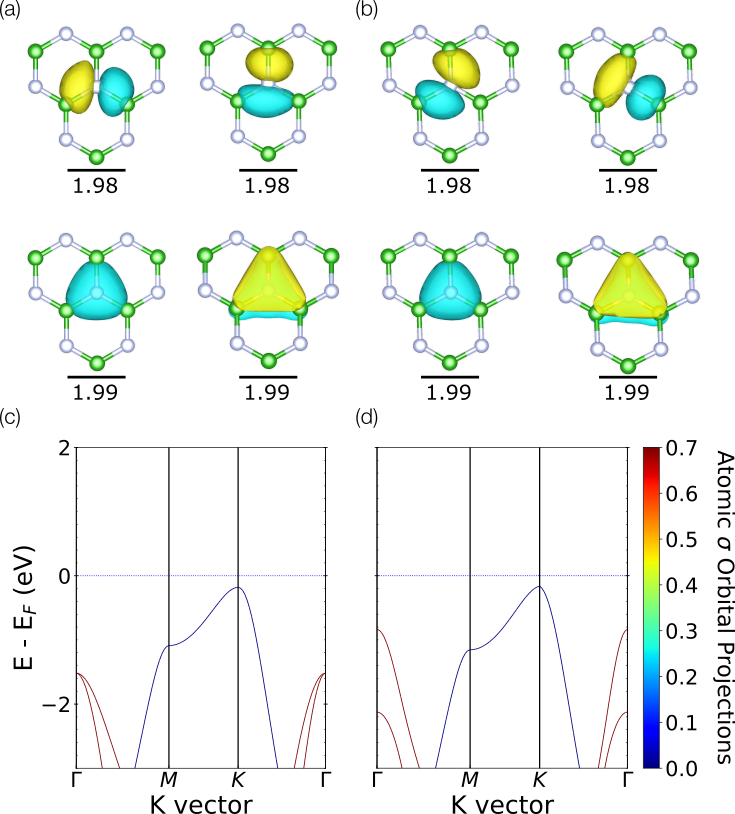}
    \caption{Bonding analysis and electronic structure of h-BN under the $E_{2g}$ phonon perturbation. Top panels display the SSAdNDP bonding motifs for the (a) unperturbed equilibrium state and (b) the distorted structure. The orbitals are sorted by ON, revealing a topologically invariant network where bonding preferences remain unaltered. Bottom panels present the electronic band structure for the (c) unperturbed and (d) perturbed systems. The color gradient represents the atomic projection onto the in-plane $\sigma$ orbitals ($s,\; p_x,\; p_y$)}
    \label{fig:hBN-4c}
\end{figure}

Consistent with this picture, our analysis identifies a 4-center (4c) configuration centered on nitrogen as the motif with the highest Occupation Number (ON) (see Fig.~\ref{fig:hBN-4c}). This strong charge localization creates a rigid electrostatic landscape that resists the topological reorganization observed in the highly covalent and delocalized B–B network of MgB$_2$. Showing that the phonon again does not change the ONs or the distribution of the bonds. This behavior, without major changes, is also obtained when a 6c configuration is chosen. 

Overall, the $E_{2g}$ phonon consistently splits electronic degeneracies in all systems studied, as evidenced by the band-structure results. However, the SSAdNDP analysis for hBN and graphene reveals that the corresponding bonds do not exhibit partially occupied or incomplete degenerate states. This indicates that, even when a phonon mode is capable of splitting electronic levels, no new bonding or antibonding configurations emerge as a result. From a Lewis-like perspective, this provides a clear explanation for why lifting degeneracies far from the Fermi surface has little or no chemical significance: a system can only respond when it possesses partially filled bands that allow the redistribution of electrons. Only in such cases—when changes in occupation numbers or bond topology occur—can the splitting be regarded as a true structural or energetic effect.

\section{\label{sec:conclusion}Conclusion}

In summary, this study demonstrates that phonon-induced band splittings can be interpreted not only from a reciprocal-space perspective but also through a chemical lens using the SSAdNDP approach. By analyzing the same $E_{2g}$ phonon mode in structurally similar yet electronically distinct systems—MgB$_2$, graphene, and hBN—we show that the lifting of electronic degeneracies does not necessarily entail a structural or energetic consequence. The SSAdNDP analysis reveals that only when the phonon perturbation affects partially filled bands near the Fermi level does a redistribution of electron density occur, reflected as changes in occupation numbers and bond topology.

The method thus provides a direct bridge between reciprocal and real space, linking the evolution of the electronic band structure with the reorganization of multicenter bonds. Moreover, the observed correlation between $\sigma/\pi$ bonding patterns and the projected orbital character of the electronic bands establishes a physically meaningful indicator of when degeneracy lifting translates into genuine chemical change.
This framework offers a chemically intuitive and transferable way to interpret phonon–electron interactions. It can be extended to other phenomena where atomic vibrations influence the electronic structure.

\section{\label{sec:four}Acknowledgements}
This work was supported by DOE BES grant DE-SC0024987 to A.N.A. J.C.E gratefully acknowledges ANID for her national doctoral scholarship year 2023, number 21231429. A.E acknowledges her FONDECYT No. 3240387. FM acknowledges support from Fondecyt grants 1231487 and 1220715, CEDENNA CIA250002, and partial support by the supercomputing infrastructure of the
NLHPC (CCSS210001)
\appendix

\section{\label{sec:Aone}Mg-Stabilized 8c Bonding Motifs}
\begin{figure}[t]
    \centering
    \includegraphics[width=1.0\linewidth]{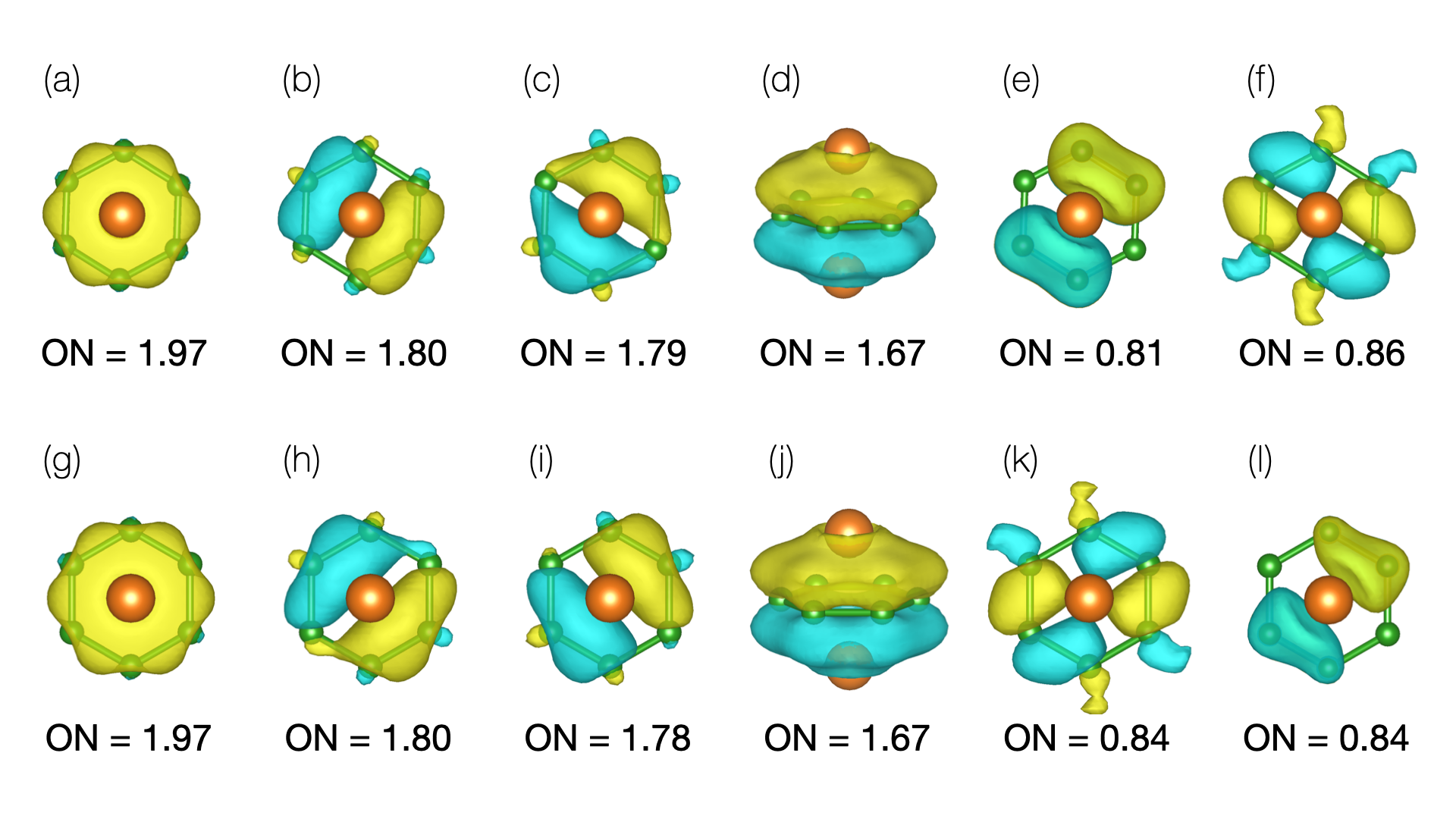}
    \caption{MgB$_2$: Bonding configurations and their occupation numbers (ON) for the 8c case. The upper panel (a-f) corresponds to the equilibrium structure, non-perturbed, while the bottom panel (g-l) shows the configuration on the frozen phonon perturbation. All the figures are in the a-b plane, except for the $\pi$ orbitals (d, j), which are in a lateral view to illustrate the nodal plane and the different phases. }
    \label{fig:8c}
\end{figure}

The inclusion of Magnesium in the 8c bonding motifs results in a increment in occupation numbers for both the delocalized $\sigma$ (\ref{fig:8c}.a) and $\pi$ (\ref{fig:8c}.d) bonding compared to the 6c configuration (Fig.\ref{fig:bands_MgB2}) This enhancement is a direct consequence of the charge donation from Mg, which the 8c framework efficiently incorporates to maximize the electronic stability of the bulk system. However, precisely because the 8c configuration represents the most stable and global contribution to the MgB$_2$ system, it is inherently robust against local distortions. Consequently, this perspective is limited in resolving the phonon-induced variations; unlike the 6c motif, the 8c representation shows no significant topological changes between the unperturbed (Fig \ref{fig:8c}.(a-f)) and perturbed (Fig \ref{fig:8c}.(g-l)) states. \\
Specifically, both the equilibrium and frozen-phonon structures in the 8c configuration exhibit an identical topology; the series begins with a highly delocalized $\sigma$ bond over the full hexagonal ring, followed by delocalized configurations across 3c and 4c centers with nearly identical ON = 1.80 |e| values, indicating a strong degeneracy of states. Then, the delocalized $\pi$ bond with ON = 1.67 |e| followed by subsequent degenerate states (2c) with ON $\approx 0.84$ |e|.The persistence of these degeneracies and ON values confirms that the global electronic backbone remains reorganized, justifying the focus on the 6c motifs in the main text to capture the symmetry-breaking effects of the phonon.
\bibliographystyle{plain}
\bibliography{bib}
\end{document}